\documentstyle[12pt,epsfig]{article}
\font\tenrm=cmr10

 1
\font\elevenrm=cmr10 scaled\magstep 1

\renewenvironment{thebibliography}[1]
 { \elevenrm
   \begin{list}{\arabic{enumi}.}
    {\usecounter{enumi}     \setlength{\parsep}{0pt}
     \setlength{\itemsep}{3pt} \settowidth{\labelwidth}{#1.}
     \sloppy
    }}{\end{list}}

\begin{document}
\begin{center}
\vglue 1cm
{\large \bf Effective Strange Quark/Antiquark Masses from\\ the Chiral Soliton Models for Baryons\\}
\vglue 0.9cm
Vladimir Kopeliovich \\
\vglue 0.3cm
Institute for Nuclear Research of RAS, Moscow; \\
Moscow Institute for Physics and Ingeneering (MIPT)
\vglue 0.1cm
\end{center}
{\rightskip=3pc
 \leftskip=3pc
 \noindent

\abstract{\baselineskip=10pt The effective strange quark and antiquark masses are estimated
from the chiral soliton approach (CSA) results for the spectrum of exotic and nonexotic baryons.
There are problems when one tries to project results of
the CSA on the simple quark models (QM): the parameter in $1/N_c$ expansion is so large for the case
of baryon spectrum that extrapolation to the real $N_c=3$ world is not possible;
rigid (as well soft) rotator model and the bound state model coincide in the first two orders
in $1/N_c$, but differ in the next orders. There is correspondence of the CSA and simple
QM predictions for pentaquarks (PQ) spectra in negative $S$ sector
 of the $\{27\}$ and $\{35\}$-plets: the effective mass of strange 
 quark is about $135 -130\,MeV$, slightly smaller for $\{35\}$.
For positive strangeness components the link between 
CSA and QM requires strong dependence of the effective $\bar{s}$ mass on 
particular $SU(3)$ multiplet. The $SU(3)$ configuration mixing is important,
it pushes the spectrum towards the simplistic model (with equal masses of the strange quark and antiquark),
and explanation of this nice property is lacking still.
The success of the CSA in describing many properties of baryons and light nuclei (hypernuclei)
 means that predictions of pentaquark states should be
considered seriously. Existence of PQ by itself is without any doubt,
although very narrow PQ may not exist. Wide, even very wide PQ should exist,
therefore, searches for PQ remain to be an actual task.}

\vglue 0.8cm}
\baselineskip=14pt
\section{Introduction}  Studies of baryon spectrum - 
nonstrange, strange, and with heavy flavors - remain to be one of main aims of accelerator physics.
Discovery of baryon states besides well established octet and decuplet, in particular,
exotic baryons, could help to the progress in understanding hadrons structure.
In the absence of the complete theory of strong interactions there are different approaches
and models; each has some advantages and certain drawbacks.
Interpretation of hadrons spectra in terms of quark models (QM) is widely
accepted, 
QM are "most successful tool for the classification and interpretation"  (R.Jaffe) of hadrons 
spectrum. The QM are to large extent phenomenological; simplicity of QM becomes a fiction
when we try to go behind e.g. 3-quark picture for baryons,
since there is no regular methods of solving relativistic many-body problem.
The number of constituents (e.g. additional $q\bar q$-pairs) is not fixed in a true
relativistic theory.

Alternative approaches, in particular, the chiral soliton approach (CSA) \cite{skyrme,wit,anw}
has certain advantages.
It is based on few principles and ingrediemts incorporated in the model lagrangian.
Baryons and baryonic systems are considered on equal footing (the look "from outside").
CSA looks like a theory, but still it is a model,
and some elements of phenomenology are present necessarily within the CSA.
It has been noted first in \cite{mp} and, for arbitrary baryon number, in \cite{vk} that
so called exotic (i.e. containing additional quark-antiquark pairs) states appear
naturally within the CSA. More definite numerical predictions for the mass of
exotic baryon with strangeness $S=+1$ were made somewhat later in \cite{hw} and
(quite definite!) in \cite {dpp}.
Results obtained within CSA for the spectrum of baryons with different values of strangeness 
mimic some features of baryons spectrum within quark models
due to Gell-Mann - Okubo relations. 

In the next section basic features and properties of the CSA are described and the Gell-Mann --- Okubo
relations for the spectrum of baryons within the rigid rotator model (RRM) of skyrmions quantization 
 are presented at arbitrary number of colors $N_c$.
Section 3 describes the bound state model (BSM) results for the baryon spectrum, the differene between the RRM 
and the BSM results is fixed, and the way to remove it is discussed. First terms of the $1/N_c$ expansion
for the effective strange quark/antiquark masses are presnted here. Section 4 contains numerical estimates of 
the effetive strange quark/antiquark masses within simple quark model where the quark/antiquark masses make additive contribution
to the baryon mass. Final section summarizes the main problems and conclusions.
\section{Features of the chiral soliton approach}
Arbitrary $SU(2)$ skyrmion is described by three profiles $\{f,\;\alpha,\;\beta\}(x,y,z)$ 
which parametrize the 
unit vector on the 3-sphere $S^3$, and the baryon number of the configuration is the degree
of the map of the 3-dimensional space $R^3$ to the 3-sphere $S^3$.
Masses, binding energies of classical configurations, moments of 
inertia $\Theta_I,\;\Theta_J$ and some other
characteristics of chiral solitons contain implicitly  
information about interaction between baryons.
Minimization of the static energy (mass) functional $M_{class}$ provides three profiles and
allows to calculate moments of inertia, etc. The details can be found in \cite{anw,vkpent,vkash,vk09}.

  The observed spectrum of states is obtained by means of the quantization procedure
and depends on the baryon quantum numbers and static characteristics of the skyrmion, 
moments of inertia $\Theta$, $\Sigma$-term ($\Gamma$), etc.
In $SU(2)$ case, the rigid 
rotator model (RRM) is most effective and successfull in describing the
properties of nucleons, $\Delta$-isobar \cite{anw}, of light nuclei \cite{msw}
 and also "symmetry energy" of nuclei with
$A< \sim 20$ \cite{ksm}.

In the $SU(3)$ case the mass formula takes place, also for RRM \cite{guad}
$$ M(p,q,Y,I,J) = M_{cl} + {K(p,q,J)\over 2\Theta_K} + {J(J+1)\over 2\Theta_\pi} +\delta M(Y,I),
\eqno (1) $$
where these terms scale as functions of the number of colors $N_c$ as $ \sim N_c,\; \sim 1,  \quad \sim N_c^{-1}$
and $ \;\sim 1,$ correspondingly; it is in fact expansion in powers of $1/N_c$.
The quantity 
$$K(p,q,J) = C_2(SU3) - J(J+1) - N_c^2B^2/12 \eqno (1a)$$
contains difference of the second order Casimir operators of the $SU(3)$ and $SU(2)$ groups,
$C_2(SU3)= (p^2+q^2+pq)/3 +p+q$, where $p,\,q$ are the numbers of upper and lower indeies in the spinor
describing the $SU(3)$ multiplet, $C_2(SU2) = J(J+1)$, $J$ being the spin of the baryon.
It is worth to mention here that formula $(1)$ takes place only for the particular case when the 
inident $SU(2)$ skyrmion is located in the $(u,d) \; SU(2)$ subgroup, i.e. it is nonstrange.
Other possibilities for the starting skyrmion configuration have been considered as well \cite{vksu3}.
Remarkable property of Eq. $(1)$ is that the total splitting of the whole $SU(3)$ multiplet is $\sim N_c$.

Mass splittings $\delta M$ are due to the term in the lagrangian
$$ {\cal L}_M \simeq - \tilde m_K^2 \Gamma {s_\nu^2\over 2}, \eqno (2)$$
$\nu$ is the angle of rotation into strange direction, 
$\tilde m_K^2 = F_K^2m_K^2/F_\pi^2 -m_\pi^2$ includes the $SU(3)$-symmetry violation in
flavor decay constants, the sigma - term $\Gamma \sim 5\, Gev^{-1} $,
moments of inertia $\Theta_\pi \sim (5 - 6)Gev^{-1},\;\Theta_K \sim (2 - 3)Gev^{-1}$, 
see \cite{vkpent,vkash} and references here. All moments of inertia $\Theta \sim N_c$.
Strange, or kaonic inertia $\Theta_K$ contains important contribution due to flavor 
symmetry breaking in meson decay constants, $F_K/F_\pi \simeq 1.23$:
$$\Theta_K = {1\over 8}\int  (1-c_f)\left[ F_K^2 + {1\over e^2}\left(f'^2 + {2s_f^2\over r^2} \right)\right] d^3r. \eqno (3) $$
This expression is valid for skyrmions originally located in the $(u,d) \;\, SU(2)$ subgroup of $SU(3)$.
"Strangeness contents" of baryons
$$ C_S = <s_\nu^2/2>_B \eqno (4) $$
can be calculated exactly with the help of the wave functions
in $SU(3)$ configuration space, for arbitrary number of colors $N_c$ \cite{vkpent, vkash}.
Some examples of the $C_S$ values at arbitrary number of colors $N_c$ are:
$$C_S("N")={2(N_c+4)\over (N_c+3)(N_c+7)}, \quad C_S("\Xi")={4\over N_c+7},
\quad C_S("\Theta")={3\over N_c+9}, \eqno (5)$$
Approximately at large $N_c$
$$C_S \simeq {2+|S|\over N_c}. \eqno (6) $$
The Gell-Mann - Okubo formula takes place in the form \cite{vkash}
$$  C_S = -A(p,q)\,Y - B(p,q) \left[ Y^2/4 - \vec I^2\right] +C(p,q), \eqno (7)$$
$ A_(p,q), B_(p,q), C_(p,q)$ depend on particular $SU(3)$ multiplet.
For the "octet", $[p,q]=[1,\,(N_c-1)/2]$,
$$ A("8") = {N_c+2 \over (N_c+3)(N_c+7)},\quad B("8") = {2 \over (N_c+3)(N_c+7)},\quad C("8") = {3 \over (N_c+7)}. \eqno (8)$$
If we try to make expansion in $1/N_c$, then the parameter is $\sim 7/N_c$.
For "decuplet" ($[p,q]=[3,\,(N_c-3)/2]$) and "antidecuplet" ($[p,q]=[0,\,(N_c+3)/2]$) the expansion parameter is $\sim 9/N_c$ and becomes worse
for greater multiplets, $"\{27\}"$-plet, $"\{35\}"$-plet, etc.
Apparently, for the realistic world with $N_C=3$ the $1/N_c$ expansion does not work and some properties
of baryon spectrum which take place at large $N_c$ may be not correct  at the realistic value $N_c=3$.
\footnote{It is instructive to note here that physics implications of the large $N_c$ extrapolation are ambiguous.
In particular, the electric charge of quarks is not fixed. If one takes the charge of the
$u (c,t)$ quark equal to $2/3$, and the charge of the $d(s,b)$ quark equal to $-1/3$, then the hypercharge
of a baryon consisting of $N_c$ quarks is integer only if $N_c/3$ is an integer.
The Gell-Mann --- Nishijima relation in this case has the form:
$$ Y= N_cB/3 +S +..., $$
see, e.g. \cite{cohen} and \cite{vkpent}.  The charges of the "proton" and neutron are $Q_p=(N_c+3)/6, \; Q_n= (N_c-3)/6$.
Another, physically attractive possibility proposed by A.Abbas \cite{abbas} is that quark charges depend on $N_c$,
$$ Q_{u(c,t)}= (1+1/N_c)/2, \qquad  Q_{d(s,b)} =(-1+1/N_c)/2.$$
These expressions follow from the Gell-Mann - Nishijima relations, $Q=I_3+Y/2$ and
$Y=B+S+C+t+b$, the latter equality is generalization of the original relation for the hypercharge,
$Y=B+S$, with flavors $C$ (charm), $b$ (beauty) ant $t$ (truth) included. $B$ is the baryon number, $B_q=1/N_c$
for quarks in the QCD with $N_c$ colors.
Evidently, the charge of any baryon is an integer number in this case \cite{abbas}. Experimental check of the quark properties
in the hypothetical (gedanken) large $N_c$ world is not possible, of course. By this reason we could guess that charges
of the quarks are
$$ Q_{u(c,t)}= (1+\alpha)/2, \qquad Q_{d(s,b)} =(-1+\alpha)/2,$$
baryon consist of $1/\alpha$ quarks, $\alpha$ being arbitrary. It would be in analogy with consideration of $D$-dimensional space-
time with noninteger $D$.}

Any chain of states connected by relation $ I= C' \pm Y/2$ reveals
linear dependence on hypercharge (strangeness), so, the CSA  mimics 
the quark model with the effective strange quark mass
$$m_S^{eff} \sim \tilde m_K^2 \Gamma [A(p,q)\mp (C'+1/2) B(p,q)], \eqno (9) $$
$C'=1$ for decuplet (antidecuplet). This is valid if the flavor symmetry breaking is included in the lowest order 
of perturbation theory.
At large $N_c$,   $\, m_S^{eff} \sim \tilde m_K^2 \Gamma/N_c,$
too much, $\sim 0.6 \,GeV$ if extrapolated to $N_c=3$. 

If we make expansion in the RRM, we obtain  for the "octet" of baryons
$$\delta M_N \simeq 2 \tilde m_K^2 {\Gamma\over N_c} \Biggl(1 - {6\over N_c} \Biggr),\qquad   
\delta M_\Lambda \simeq \bar m_K^2 {\Gamma\over N_c}\Biggl(3 - {21\over N_c} \Biggr) $$
$$\delta M_\Sigma \simeq \bar m_K^2 {\Gamma \over N_c}\Biggl(3 -{17\over N_c} \Biggr), \qquad
\delta M_\Xi \simeq \tilde m_K^2 {\Gamma \over N_c}\Biggl(4 - {28\over N_c} \Biggr), \eqno (10) $$

For the "decuplet" of baryons we have after such expansion
$$\delta M_\Delta \simeq 2 \tilde m_K^2 {\Gamma\over N_c} \Biggl(1 - {6\over N_c} \Biggr), \quad ... \qquad   
\delta M_\Omega \simeq \bar m_K^2 {\Gamma\over N_c}\Biggl(5 - {45\over N_c} \Biggr), \eqno (11) $$
equidistantly for all components.
For positive strangeness components of exotic multiplets we obtain from our previous results \cite{vkash,vk09}
$$\delta M_{\Theta_0, J=1/2} \simeq  \tilde m_K^2 {\Gamma\over N_c} \Biggl(3 - {27\over N_c} \Biggr),\qquad   
\delta M_{\Theta_1,J=3/2} \simeq \bar m_K^2 {\Gamma\over N_c}\Biggl(3 - {25\over N_c} \Biggr), $$
$$\delta M_{\Theta_2, J=5/2} \simeq \tilde m_K^2 {\Gamma \over N_c}\Biggl(3 - {23\over N_c} \Biggr), \eqno (12) $$

These results are summarized in Table 1.
\section{The bound state model} Within the bound state model (BSM) 
antikaon field is bound by the $SU(2)$ skyrmion \cite{ckr,wk}. The mass formula takes place
$$ M = M_{cl} + \omega_S + \omega_{\bar S} + |S| \omega_S + \Delta M_{HFS} \eqno (13) $$
where strangeness and antistrangeness excitation energies 
$$\omega_S= N_c(\mu-1)/8\Theta_K,\;\;\omega_{\bar S}= N_c(\mu+1)/8\Theta_K, \eqno (14) $$
$$\mu = \sqrt{1+\bar m_K^2/M_0^2}\simeq 1+{\bar m_K^2\over 2M_0^2},\quad
 M_0^2=N_c^2/(16\Gamma\Theta_K)\sim N_c^0,\quad \mu \sim N_c^0. \eqno (15)$$ 

The hyperfine splitting correction depending on hyperfine splitting
constants $c$ and $\bar c$, and "strange isospin" $I_S=|S|/2$ equals \cite{wk} 
$$\Delta M_{HFS}(S,I,J) = {J(J+1)\over 2\Theta_\pi} 
+\frac{(c_S-1)[J(J+1)-I(I+1)] +(\bar c_S-c_S)
I_S(I_S+1)}{2\Theta_\pi} \eqno (16) $$
and is small at large $N_c$, $\sim 1/N_c$, and for heavy flavors (see \cite{wk,vkpent,vk09} where the hyperfine
splitting constants $c_S,\; c_{\bar S}$ are presented).
For anti-flavor (positive strangeness) certain changes should be done:
$\omega_S \to \omega_{\bar S}$  and $c_S \to c_{\bar S}$ in the last term.
The baryon states in the BSM are labeled by their strangeness (flavor in general case), isospin and spin, but do 
not belong apriori to a definite $SU(3)$ multiplet $[p,q]$, and can be some mixture of different $SU(3)$ multiplets.

In this way we obtain for the "octet" \cite{vkpent}
$$\delta M_N =2 \tilde m_K^2 {\Gamma\over N_c}, \quad ... \quad\delta M_\Xi = \tilde m_K^2 {\Gamma \over N_c}\Biggl(4 - {4\over N_c} \Biggr), \eqno (17)$$
and for "decuplet"
$$\delta M_\Delta \simeq 2 \tilde m_K^2 {\Gamma\over N_c} , \quad ... \quad   
\delta M_\Omega \simeq \bar m_K^2 {\Gamma\over N_c}\Biggl(5 - {15\over N_c} \Biggr), \eqno (18) $$
Total splitting of the "octet" and "decuplet"
$$ \Delta_{tot}("8",BSM) = \tilde m_K^2 {\Gamma \over N_c}\Biggl(2 - {4\over N_c} \Biggr), \;
 \Delta_{tot}("10",BSM) = \tilde m_K^2 {\Gamma \over N_c}\Biggl(3 - {15\over N_c} \Biggr).  \eqno (19)$$    
In the BSM the mass splittings are bigger than in the RRM.

For exotic $S=+1$ $\Theta$- hyperons we obtain in BSM \cite{vkpent,vkash,vk09}
$$\delta M^{BSM}_{\Theta_0,J=1/2}=  \bar m_K^2\Gamma\Biggl({3\over N_c}-{9\over N_c^2}\Biggr), \quad 
\delta M^{BSM}_{\Theta_1,J=3/2}= \bar m_K^2\Gamma\Biggl({3\over N_c}-{7\over N_c^2}\Biggr), $$
$$  \delta M^{BSM}_{\Theta_2,J=5/2}= \bar m_K^2\Gamma\Biggl({3\over N_c}-{5\over N_c^2}\Biggr)\eqno (20)$$
and again considerable difference from the RRM results presented in Eq. $(12)$ takes place.

From the comparison of these results with previous section we conclude that the RRM used for 
prediction of pentaquarks in \cite{mp,hw,hwvk} is different from  
the BSM model, used in the paper \cite{ikor} to disavow the $\Theta^+$.

\begin{center}
\begin{tabular}{|l|l|l|l|l|l|}
\hline
&$\quad \{8\}$&\quad $\{10\}$ &$\quad \{\overline{10}\}$ &$\quad \{27\} $& $\quad \{35\}$\\
\hline
$m_s^{RRM}$&$ 1-8/N_c$ &$1- 11/N_c$&$\quad  -  $&$1 - 18/N_c $ &$1 - 56/3N_c $ \\
\hline
$m_s^{BSM}$&$ 1-2/N_c$ &$1-5/N_c$&$\quad  - $&$\quad  - $ &$\quad  - $   \\
\hline
\hline
$m_{\bar s}^{RRM}$&$\quad -$ &$\quad - $&$ 1- 15/N_c$&$1-13/N_c$ &$1-11/N_c$ \\
\hline
$m_{\bar s}^{BSM}$&$\quad  - $ &$\quad  -$&$ 1- 9/N_c$&$1-7/N_c$ &$1-5/N_c$ \\
\hline
\end{tabular}
\end{center}

{\rightskip=3pc
 \leftskip=3pc
\noindent 
{\elevenrm {\bf Table 1.}  First terms of the $1/N_c$ expansion for the effective strange quark 
mass (the upper two lines) and the antiquark mass (the lower two lines) within different $SU(3)$ multiplets, 
in units $\bar m_K^2 \Gamma/N_c$. Empty spaces are left in the cases of theoretical uncertainty. The assumption concerning strange 
quarks/antiquarks sea should be kept in mind, see explanation in the text.}\\

{\rightskip=-0.01pc
 \leftskip=-0.01pc

The mass $m_s$ for the "octet" is defined as half of the splitting between
nucleon ($Y=1,\;I=1/2$) and $\Xi$-hyperon ($Y=-1,\;I=1/2$). For "decuplet" $m_s$ is defined as
$1/3$ of the splitting between the isobar, $Y=1,\; I=3/2$ and the  $\Omega$-hyperon, $Y=-2,\;I=0$.
For higher multiplets the strange quark masses are obtained from the negative strangeness
sectors, as differences of masses of states with $(Y,I) = (0,2)$ and $(-1, 3/2)$, or $(-1, 3/2)$ and $(-2, 1)$
for the $"\{27\}"$-plet, and states with $(Y,I)=(1,5/2)$ and $(0,2)$, or $(0,2)$ and $(-1,3/2)$, etc. for
the $"\{35\}"$-plet.

To define the masses of the strange antiquarks we assumed first that the strange quarks sea within
exotic multiplets is the same as in the "octet" and "decuplet", i.e.
$$C_S^{RRM}(sea)= C_S^{RRM}("8",sea)\simeq C_S^{RRM}("10",sea) \simeq {2\over N_c} \left(1-{6\over N_c}\right), \eqno (21)$$ 
and similar for the BSM:
$$C_S^{BSM}(sea)= C_S^{BSM}("8",sea)\simeq C_S^{BSM}("10",sea) \simeq {2\over N_c} , \eqno (22)$$ 
Strangeness contents of the nucleon and delta-isobar coincide in the leading and next-to-leading orders
of the $1/N_c$ expansion.
Then, the strange antiquark mass equals
$$m_{\bar s}(B) = \bar m_K^2\Gamma \left[C_S(B) -  C_S(sea)\right], \eqno (23)$$
where $B$ is the exotic, strangeness $S=+1$ baryon. These assumptions lead to the results presented
in Table 1.

It follows from Table 1 that 
the addition of the term to the BSM result, possible due to normal ordering 
ambiguity present in the BSM (I.Klebanov, VBK, 2005, unpublished)
$$ \Delta M_{BSM} = -6\bar m_K^2 {\Gamma\over N_c^2}(2+|S|) \eqno (24)$$
brings results of the RRM and BSM in agreement - for nonexotic and exotic states.
This procedure looks not quite satisfactorily: if we believe to the RRM, there is no need 
to consider the BSM and to bring it in correspondence with the RRM.
Anyway, the RRM and the BSM in its accepted form are {\it different models}.

The $SU(3)$ configuration mixing of exotic baryon multiplets has been studied in \cite{hwvk},
but was ignored in most of previous papers devoted to exotic baryons. 
For antidecuplet mixing with nonexotic components of the octet
is important, it decreases slightly the total splitting,
and pushes $N^*$ and $\Sigma^*$ toward higher energy. Apparent contradiction with simplest assumption of equality of masses
of strange quarks and antiquarks $m(s) = m(\bar s)$ takes place.

For decuplet configuration mixing with components of the $\{27\}$-plet increases total splitting of the decuplet 
considerably, but
approximate  equidistancy in the position of the decuplet components still remains. 
The configuration mixing should be included within the QM as well: states with different numbers of $q\bar q$ pairs can mix, and 
this is complicated not resolved problem, see next section. 

The rotation-vibration approach (RVA) by H.Weigel and H.Walliser \cite{hwhw}
unifies the RRM and BSM in some way,  $\Theta^+$ has been confirmed with somewhat higher
energy and considerable width $(\Gamma_\Theta \sim 50\,MeV)$.
\section{Comparison with simple quark models}  It is possible to make comparison of the CSA results with expectations
from simple quark model in pentaquark approximation, or $N_c+2$ approximation for arbitrary $N_c$ (projection of CSA on QM).
The masses $m_s$, $m_{\bar s}$ and $m(s\bar s)$ come into play.
The following diquark-diquark-antiquark wave functions are considered usually. For antidecuplet
$$ |\overline{10}, Y=2, I=0, I_3=0> \, (\Theta^+_0) \sim  \bar s [du][du]; $$
$$ |\overline{10}, 1, I=1/2,I_3=1/2> \sim -\bar d [du][du] + \bar s [su][du] + \bar s [du][su]; $$
$$ |\overline{10}, 0, I=1,I_3=1> \sim -\bar d [su][du] + \bar d [du][su] + \bar s [su][su]; $$
$$ |\overline{10}, -1, I=3/2,I_3=3/2> \sim \Xi^*_{3/2} \sim -\bar d [su][su] \eqno (25). $$
$[q_1q_2]$ means antitriplet in color, singlet in spin, antisymmetric in flavor combination (antitriplet) considered in
\cite{jw}, see also \cite{close}, and called "good" diquark, according to \cite{fw}. These wave functions can be 
obtained by action of the $U$-spin operator, $Ud=s, $  
$U\bar s = -\bar d$, and should be normalized properly. 
For each of the isomultiplets the states with other possible 3-d projection
of the isospin $I_3$ can be obtained from the wave function of the highest state with $I_3=I$ by
action of the lowering operator $I^-$.
 Obviously, the weight of the $s\bar s$ pair
is $0,\; 2/3;\; 1/3,\;$ and $0$ in these 4 states of the antidecuplet.

For $\{27\}$-plet we have
$$ |27, Y=2, I=1,I_3=1> \, (\Theta^{++}_1) \sim \bar s (uu)[du] + \bar s (du)(uu); $$
$$ |27, 1, I=3/2, I_3=3/2> \sim -\bar d (uu)[du] + \bar s (uu)[su]; $$
$$ |27, 0, I=2, I_3=2> \sim -\bar d (uu)[su]; $$
$$ |27, -1, I=3/2,I_3=3/2> \sim -\bar d (su)[su]; $$
$$ |27, -2, I=1, I_3=1> \sim -\bar d (ss)[su] \eqno (26) $$
The weight of the $s\bar s$ pair is $0,\; 1/2,\; 0,\; 0$ and $0$.
Here $(q_1q_2)$ is triplet in spin, symmetric in flavor (i.e. 6-plet in flavor) diquark
configuration, so called "bad" diquark (in color it is also antitriplet, similar to "good" diquark).

For the $\{35\}$-plet we have, also for the components with maximal isospin
$$ |35, Y=2, I=2, I_3=2> (\Theta^{+++}_2) \sim \bar s (uu)(uu); $$
$$ |35, Y=1, I=5/2, I_3=5/2> \sim \bar d (uu)(uu); $$
$$ |35, Y=0, I=2, I_3=2> \sim \bar d (su)(uu)+ \bar d (uu)(su); $$
$$ |35, Y=-1, I=3/2, I_3=3/2> \sim 2\bar d (su)(su)+ \bar d (uu)(ss)+\bar d (ss)(uu); $$
$$ |35, Y=-2, I=1, I_3=1> \sim \bar d (su)(ss)+ \bar d (ss)(su); $$
$$ |35, Y=-3, I=1/2, I_3=1/2> \sim \bar d (ss)(ss). \eqno (27) $$
There are no $s\bar s$ pairs in the QM wave functions of the $35$-plet components. 
Table 2, given also previously in \cite{vkpent,vk09}, summarizes these results, in comparison with
the rigid rotator model calculations.
The upper lines (for each of the $SU(3)$ multiplets) in the table show the contribution of 
masses of the strange quark $m_s$, antiquark $m_{\bar s}$
and the quark-antiquark pair $m_{s \bar s}$ into the mass of the quantized state in the simple
quark model where the quark mass makes additive contribution to the mass of the baryon, according to wave functions
given ib Eq. (25)-(27).
E.g., for the antidecuplet it is $m_{\bar s}$, $2 m_{s\bar s}/3$, $m_s+m_{s\bar s}/3$ and $2 m_s$ (the number of colors
$N_c=3$).
Here we keep different masses for the strange quark and antiquark, and $m_{\bar s s}$ may be different from the sum
$m_{\bar s} + m_s$. However, without configuration mixing simple relations take place for the mass of the strange
quark-antiquark pair $m_{s \bar s}$:
$$m_{s \bar s}|_{  \{\overline{10}\}} =m_s|_{\{\overline{10}\}}+m_{\bar s}|_{\{\overline{10}\}},  \eqno (28) $$
and
$$m_{s \bar s}|_{(\{27\}} =m_s|_{\{27\}} +m_{\bar s}|_{\{27\}}.  \eqno (29) $$
These equalities are in fact the consequences of the Gell-Mann --- Okubo relations for the masses of the
$SU(3)$ multiplet components.

The expressions $(25) \,-\,(27)$ for the wave functions will be more complicated if the number of colors $N_c>3$, and the
$N_c+2$ approximation should be investigated instead of the pentaquark approximation.
The contribution of the $m_{s\bar s}$ to the masses of states with hypercharge $Y=1$, or zero strangeness will 
be   $(N_c+1)/(N_c+3)$ for "antidecuplet", $(N_c-1)/(N_c+1)$ for the "$\{27\}$"-plet, and $(N_c-3)/(N_c-1)$ for "$\{35\}$"-plet,
which obviously go over to $2/3$, $1/2$ and $0$, shown in Table 2.

Configuration mixing leads to modification and complications of these contributions, expressions for the wave functions
(25) --- (27) do not hold, and our estimates of the strange quark/antiquark masses become very approximate.
The numerical values of the mass differences of the baryon state and the nucleon, calculated in
the rigid rotator model of the CSA \cite{hwvk}, are shown in the 
next lines. Besides the mentioned strange quark/antiquark masses these differences contain the differences
of the rotation energies depending on the moments of inertia.
Simple relations can be obtained from this table for the effective $s-$quark/antiquark masses, since in the differences
of masses of states which belong to the same $SU(3)$ multiplet the rotational energies cancel, according to 
previously given in \cite{vkpent,vk09} expressions, as well as the strange quarks sea contributions.
From the total splitting of antidecuplet we obtain, before the $SU(3)$ configurations mixing
$$[2m_s - m_{\bar s}]_{\overline{10}} = \bar m_K^2\Gamma/8\simeq 272\,Mev. \eqno (30)$$
Configuration mixing decreases this quantity to $247\,MeV $.

{\tenrm
\begin{center}
\tenrm
\begin{tabular}{|l|l|l|l|l|l|}
\hline
$|\overline{10},2,0>$&$|\overline{10},1,{1\over 2}>$&$|\overline{10},0,1>$&
$|\overline{10},-1,{3\over 2}>$& &\\
\hline
$m_{\bar s}+... $         &$2 m_{s\bar s}/3+... $   &$m_s+m_{s\bar s}/3+... $ & $2 m_s+...$& & \\
\hline
564 & 655 & 745 & 836 & &\\
600 & 722 & 825 & 847 & & \\
\hline
\hline
$|27,2,1>$&$|27,1,{3\over 2}>$&$|27,0,2>$&$|27,-1,{3\over 2}>$&$|27,-2,1>$&\\
\hline
$m_{\bar s}+...$         &$ m_{s\bar s}/2+...$   &$ m_s+...$&$ 2 m_s+...$&$ 3m_s+...$  & \\

\hline
733 & 753 & 772 & 889 & 1005 &\\
749 & 887 & 779 & 911 & 1048 &\\
\hline
\hline
$|35,2,2>$&$|35,1,{5\over 2}>$&$|35,0,2>$&$|35,-1,{3\over 2}>$&$|35,-2,1>$&$|35,-3,{1\over 2}>$ \\
\hline
$ m_{\bar s}+...$   &$ ...  $   &$ m_s+...$&$2 m_s+...$&$3m_s+...$ &$ 4 m_s+...$ \\
\hline
1152 & 857 & 971 & 1084 & 1197 & 1311 \\
1122 & 853 & 979 & 1107 & 1236 & 1367 \\
\hline
\end{tabular}
\end{center} }
{\rightskip=3pc
\leftskip=3pc
\noindent 
{\elevenrm {\bf Table 2.} The strange quark (antiquark) masses contributions to the masses of baryons
according to the simple wave functions in the pentaquark approximation, $N_c=3$ (first lines for each of
exotic $SU(3)$ multiplets of baryons). 
Numerical values, in $MeV$, are given for the mass differences of the baryon state and the nucleon, 
second and 3-d lines for each of multiplets, within the CSA. 2-d lines - without configuration 
mixing, and 3-d line - configuration mixing included. $\Theta_K = 2.84 \;GeV^{-1}$, $\Theta_{\pi} = 5.61\; GeV^{-1}$ and
$\Gamma = 1.45 \;GeV^{-1} $ \cite{hwvk}.} \\

{\rightskip=-0.001pc
 \leftskip=-0.001pc

In the simplistic model ($m_{\bar s} = m_s = m_{\bar s  s}/2$) we would obtain $\Delta (\{ \overline{10}\}) = m_s$,
in contradiction with numerical value from the CSA. Remarkably, that configuration mixing decreases the splitting 
of antidecuplet ($247 \,Mev$ instead of $272\,Mev$), thus  pushing the result of the CSA toward the simplistic QM.
If we believe that the strange quark mass within the antidecuplet is in usually accepted interval, $\sim 120 - 150 \,Mev$,
then the strange antiquark mass should be unusually small, even negative.

From splittings within $\{27\}$-plet  we obtain, before mixing
$$[m_s - m_{\bar s}]_{27}= \bar m_K^2\Gamma/56 =\;39\,Mev, \eqno (31)$$
Configuration mixing decreases this to $30\,Mev$.
So, strange antiquark is lighter than strange quark, and again the configuration mixing pushes the effective strange antiquark
mass towards the simplistic model.
From the negative strangeness sector of the $\{27\}$-plet we get 
$$[m_s]_{27}= 3\bar m_K^2\Gamma/56 \simeq 117\,Mev, \eqno (32)$$
before mixing and $[m_s]_{27}\simeq 135 Mev$ after the configuration mixing, 
in reasonable agreement with accepted value of the strange quark mass.
The mass of the cryptoexotic component of the $27$-plet $S=0,\; I=3/2$ increases considerably due to strong
mixing with the corresponding component of the decuplet.

\begin{center}
\begin{tabular}{|l|l|l|l|l|l|}
\hline
&$\;\; \{8\}$& $\;\{10\}$ &$\; \{\overline{10}\}$ &$\; \{27\} $& $\; \{35\}$\\
\hline
$m_s^{RRM}(no\; mix)$&$1/12   $ &$\;1/24  $&$\;  -  $&$\; $3/56$ $ &$\; 5/96 $ \\
\hline
$m_{\bar s}^{RRM}(no\;mix)$&$\; -$ &$\; - $&$\; -  $&$\;2/56$ &$ 13/96$ \\
\hline
$(2m_s - m_{\bar s})^{RRM}(no\;mix)$&$\; -$ &$\; - $&$1/8   $&$\;1/14$ &$-1/32$ \\
\hline
\end{tabular}
\end{center}
{\rightskip=3pc
 \leftskip=3pc
{\elevenrm{\bf Table 3.}  The values of the strange quark/antiquark masses in the RRM without configuration mixing 
(no mix), in units $\bar m_K^2\Gamma$ in the realistic case  $N_c=3$. The strangeness contents $C_S$
of the states used for this evaluation are taken from \cite{vkash,vk09}.}\\

{\rightskip=-0.01pc
 \leftskip=-0.01pc

From the $\{35\}$-plet  the difference of energies of states with $S=+1$ and $S=0$ gives
the mass of the strange antiquark
$$ [m_{\bar s}]_{35}= 13\bar m_K^2\Gamma/96 \simeq 295\,Mev. \eqno (33)$$
Configuration mixing deceases this quantity down to $\simeq 270 \,Mev$.
From the negative strangeness sector of the $35$-plet we obtain
$$ m_s|_{35} = 5\bar m_K^2\Gamma/96\simeq 113\,Mev,  \eqno (34)$$
and the configuration mixing increases this value up to $\sim 130\,Mev$, again
in reasonable agreement with commonly accepted value.

Our results for the strange quark, antiquark masses and the combination $2m_s-m_{\bar s}$ are presented
in Table 3 (in units of $\overline m_K^2\Gamma$) and in Table 4 (numerically, in $MtV$).
Strong dependence of the $s$-antiquark mass on the multiplet takes place, in qualitative agreement
with the $1/N_c$ expansion, presented in Table 1.
It is a challenge to theory to understand, is it an artefact of CSA, or has physical 
meaning.

For the octet the quark mass $m_s$ is the half of the total mass splitting, for decuplet it is $1/3$ 
of the decuplet splitting, for higher multiplets $m_s$ is defined by splittings in the negative strangeness
sector, components with largest isospin.

\begin{center}
\begin{tabular}{|l|l|l|l|l|l|}
\hline
&$\; \{8\}$& $\{10\}$ &$\; \{\overline{10}\}$ &$\; \{27\} $& $\; \{35\}$\\
\hline
$m_s^{RRM}(no\, mix)$&$ 181   $ &$ 91  $&$\;  -  $& $\,117 $  &$\,113 $ \\
\hline
$m_s^{RRM}(mix)$&$ 196   $ &$ 134  $  & $\;  -  $& $\,135 $  &$\,130 $ \\
\hline
\hline
$m_{\bar s}^{RRM}(no\, mix)$&$\; -$ &$\; - $&$\; -   $&$\; 78 $ &$\, 295$ \\
\hline
$m_{\bar s}^{RRM}(mix)$&$\; -$ &$\; - $&$\; -   $&$ \,105 $ &$ \,270$ \\
\hline
\hline
$(2m_s - m_{\bar s})^{RRM}(no\, mix)$&$\; -$ &$\; - $&$ 272 $&$\,156 $ &$-69$ \\
\hline
$(2m_s - m_{\bar s})^{RRM}( mix)$&$\; -$ &$\; - $&$ 247   $&$\,165 $ &$-10$ \\
\hline

\end{tabular}
\end{center}
{\rightskip=3pc
 \leftskip=3pc
{\elevenrm{\bf Table 4.}  Numerical values of the effective strange quark/antiquark masses (in $MeV$) 
in the RRM without configuration mixing (no mix), and with configuration mixing (mix). 
The results of previous papers \cite{vkpent,vkash,vk09} are used here. $\Theta_K = 2.84 \;GeV^{-1}$, $\Theta_{\pi} = 5.61\; GeV^{-1}$ and
$\Gamma = 1.45 \; GeV^{-1} $ \cite{hwvk}.}\\

{\rightskip=-0.01pc
 \leftskip=-0.01pc

Strong dependence of the strange antiquark mass and the combination  $2m_s - m_{\bar s}$ on the $SU(3)$ multiplet
becomes somewhat softer after inclusion of the configuration mixing, according to Table 4.
It should be kept in mind, however, that the above presented wave functions for the pentaquarks, Eq. $(25 - 27)$
do not hold after configuration mixing, and the definition of the quark masses itself becomes more shaky
in this case.
\section{Problems and conclusions}
 We conclude with following remarks and statements. 
The parameter in $1/N_c$ expansion is large for the case of the baryon spectrum, extrapolation to real world
is not possible in this way.
Rigid (soft as well) rotator model and the bound state model coincide in
the first two orders of $1/N_c$, but differ in the next orders.
 Configuration mixing is important, as it follows from the RRM results.   

 There is correspondence of the chiral soliton model (RRM) and
the quark model predictions for pentaquarks spectra in the negative strangeness sector
of the $\{27\}$ and $\{35\}$-plets: the effective mass of strange 
 quark is about $135 -130\,MeV$, slightly smaller for $\{35\}$. Our estimates are in reasonable  agreement with
calculations made from different points, see e.g. \cite{kkp} where the effective strange quark mass is ontained
from analysis of the $m_s^2$-corrections to Cabibbo-suppressed tau lepton decays within the perturbative QCD dynamics.

 For positive strangeness components the link between the
CSA and QM requires strong dependence of the effective $\bar{s}$ mass on 
particular $SU(3)$ multiplet. Configuration mixing slightly pushes spectra towards 
the simplistic model with equal masses of strange quark and antiquark, 
but explanation of this remarkable consequence of the configuration mixing is absent still.

In spite of these problems, the chiral soliton models, based on few principles and
ingredients incorporated in the effective lagrangian, allow to describe
qualitatively, in some cases even quantitatively, various chracteristics 
of baryons and nuclei --- from ordinary $(S=0)$ nuclei to known hypernuclei.
 This suggests that predictions of pentaquark states should be
considered seriously. Existence of PQ by itself is without any doubt,
although very narrow PQ may not exist. Wide, even very wide PQ should exist.

 In view of existing theoretical uncertainties, further experimental investigations
of baryon spectrum, in particular, searches for exotic baryons - strange, charmed or beautifal, 
wide or narrow - are of great importance.\\

\end{document}